\documentstyle[12pt]{article}
\author{$I.V.Ostrovskii^{\dagger}, S.V.Saiko^{\dagger}, O.Ya.Olikh^{\dagger}, H.G.Walther^{\ddagger}$}
\title{ACOUSTOELECTRIC STUDY OF INTERFACE TRAPPING DEFECTS IN GaAs EPITAXIAL STRUCTURES}
\begin{document}

\maketitle

\begin{center}
$^{\dagger}$ Kiev Shevchenko University, Physics Faculty, Kiev - 252022, Ukraine

${\ddagger}$ Friedrich Schiller University, Institute for Optics and Quantum Electronics, 07743 Jena, 
Germany
\end{center}

\begin{abstract}
A new acousto-electrical method making use of transient transverse acoustoelectric 
voltage (TAV) to study solid state structures is reported. This voltage arises after a surface 
acoustic wave (SAW) generating the signal is switched off. Related measurements consist in 
detecting the shape of transient voltage and its spectral and temperature dependence. Both theory 
and experiment show that this method is an effective tool to characterize trapping centers in the 
bulk as well as at surfaces or interfaces of epitaxial semiconductor structures.


PACS number(s): 78.70; S7.1
\end{abstract}

\begin{center}
\section{INTRODUCTION}
\end{center}

Different methods are known to study crystal defects and deep levels in 
semiconductors in solid state physics. Most of them, including modern Scanning 
Tunneling Microscopy techniques, allow to characterize surface defects. At the same time 
multilayer and epitaxial semiconductor structures are strongly influenced by defects [1-3] 
which can be located at the interfaces between epilayers and substrate. Up to now no 
appropriate experimental methods exist to characterize these interface defects. Recently 
some publications reported the transient acoustoelectric effect in semiconductors with 
defects [4-6]. Transient transverse acoustoelectric voltage (TAV) generated in 
semiconductors has already been used in the past [7,8], but this using hasn't be aimed to 
interface defect characterization and hasn't include corresponding acousto-optical 
measurements.

If the TAV method is applying some difficulties of interpretation of the 
experimental results have to be surmounted. The point is that the TAV amplitude 
strongly depends on several parameters, among them the influence of piezoelectric field 
strength $E_v$ on local centers has to be mentioned. To avoid these difficulties we propose to 
measure the transient TAV signal just after the excitation pulse has be switched off. 

\begin{center}
\section{THEORY}
\end{center}

\begin{center}
\subsection{MECHANISMS OF THE TAV ORIGIN}
\end{center}

TAV is one of the manifestations of ultrasonically activated redistribution of the 
electrical charges in semiconductors. It is generated across layered systems consisting of 
piezoelectric and semiconducting materials due to a piezoactive surface acoustic wave 
(SAW) propagating along the piezoelectric. The SAW piezoelectric field penetrates 
inside the semiconductor causing charge carriers redistribution in the near surface region.

Two main mechanisms of generating the TAV effect exist in semiconductors. The 
first one, so-called "concentration-effect" is due to the variable component of sample 
conductivity ${\sigma_v}$ [9]. It is a result of electron concentration redistribution under the action 
of the variable piezoelectric field $E_v$ of SAW. Hence, the direct component of 
acoustoelectric current is defined by:

\begin{equation}\label{1}
j_0=\overline{(\sigma_vE_v)}^{T_s}                         
\end{equation}                                  

where the averaging is carried out on acoustic wave period $T_s$. The relaxation time of this 
TAV signal component is defined by the Maxwell's relaxation time $\tau_m$ of free charge 
carriers $(\tau_m=\epsilon\epsilon_0/\sigma_0)$. Usually $\tau_m$ is much less than the period length of SAW. Therefore 
one can suppose that the relaxation time of the "concentration" TAV is equals to zero.

The second mechanism of the TAV generation is connected with semiconductor 
defects. The presence of the high-frequency SAW electrical field in the near surface 
region results an increasing of the free charge carriers concentration. This, in turn, causes 
an increasing of the electrical charge captured on deep trapping levels, which are located 
either at interface or surface. As a result a direct electrical field perpendicularly to the 
sample surface is occurring. The amplitude of this "trap" TAV component is proportional 
to the excess concentration of the charge captured on trap levels [10].

Let us assume that the concentration of this charge is ${\Delta}n_t(t)$. For the simplest case 
of only one type of surface defects existing the transient TAV signal can be written as 

\begin{equation}\label{2}
V_{ae}(t)=C_{ae}{\Delta}n_t(t)                         
\end{equation} 
                                                   
where the coefficient $C_{ae}$ depends on sample parameters.

The voltage $V_{ae}(t)$ is connected to the relaxation of the electrical nonequilibrum 
charge trapped by surface defects. As far as the relaxation time does not depend on the 
particular mechanism of excitation of the electronic subsystem, it is possible to use 
standard theoretical approach. The relaxation time $\tau$ is a function of trap level parameters. 
In the following we will consider an n-type semiconductor. Then it holds

\begin{equation}\label{3}
\tau=\frac{1}{N_cV_TS_n}e^{E_t/kT}                         
\end{equation} 

where $N_c$ is the density of states of the conductivity band, $V_T$ is a thermal velocity of free 
electrons, k is Boltzmann's constant, T is temperature, $E_t$ is an energy depth of the 
electronic trap level counted from the edge of conductivity band and $S_n$ is an effective 
cross section of electron capture by trap centers. 

Usually different types of trap levels exist. In this case the excess concentration ${\Delta}n_{ti}$ 
of  charge carriers captured by i - type surface traps, can be written as:

\begin{equation}\label{4}
\frac{d{\Delta}n_{ti}}{dt}=-\frac{{\Delta}n_{ti}}{{\tau}_i}+F_i(t)                         
\end{equation} 
                                           
where $\tau_i$ represents the characteristic relaxation time of i-type level, $F_i(t)$ is an external 
force, which initiates the capturing by surface traps. The physical nature of said force is 
connected to the piezoelectric field of SAW.

If the ultrasonic wave amplitude is modulated by rectangular pulses the external 
force F(t) has the form

\begin{eqnarray}\label{5}
{F(t)}=
\biggl{\lbrace} \begin{array}{cc}
const,& 0<t<T_p\\
0, &T_p<t
\end{array}               
\end{eqnarray} 
                          
where $T_p$ is the pulse duration. The radio frequency impulses and corresponding TAV 
pulse are presented on Fig.1. The TAV pulse shows a monotonous increase ("AB") and a 
monotonous decrease ("BC"). This shape of the TAV signal is typical for the trap 
component. The measurement of the shape of the transient TAV signal is made along 
"BC" in Fig.1. Note that in this case the external force vanishes (F(t)=0) during our 
measurements. In this case the solution of the equ.4 is a decreasing exponential curve 
with relaxation time ${\tau}_i$. For different trap levels in the sample the TAV signal can be 
expressed by a sum

\begin{equation}\label{6}
V_{ae}(t)=V_C+\sum_{i=1}^{N}V_ie^{-t/{\tau_i}}                         
\end{equation} 

where the times $\tau_i$ correspond to various i - levels. The summation has to be carried out 
over all existing traps in the near surface region. The coefficients $V_i$ are proportional to 
the concentrations of the various trap types. The first term in equ.6 presents the 
contribution in TAV of "very slow" levels. Such levels can exist in $A_2B_6$ - compounds or 
in MIS-structures with relaxation times $\tau$ of about some hours while modern $A_3B_5$ 
epilayer structures do not have such slow levels. The signs of $V_C$ and $V_i$ are opposite for 
electron and hole capture centers. 

Thus, it is possible to determine the relaxation time $\tau$ by measuring the TAV 
signal. The transient TAV relaxation time $\tau$ is determined by the rate of thermal emission 
and trapping of charge carriers on surface levels. In turn, the rate of reaching the 
thermodynamic balance between capture centers and conduction zones in a 
semiconductor structure depends on characteristic parameters of these centers, such as 
energy depth $E_t$ and effective cross section $S_n$. The calculation of $S_n$ by means of equ.3 is 
successful only if the energy levels $E_{ti}$ is known. For this reason additional information 
about  $E_{ti}$ and relaxation time $\tau_i$ is needed. It can be obtained by measuring the optical 
spectra of TAV signals or the thermal dependence of  TAV signal.

\begin{center}
\subsection{ANALYSIS OF TAV SPECTRAL DEPENDENCIES}
\end{center}

The illumination of a semiconductor surface changes of a distribution of free and 
captured charge carriers. Now we consider the influence of monochromatic illumination 
on various components of the TAV signal. For singly charged electron traps the process 
of electron capturing follows the equation

\begin{equation}\label{7}
\frac{dn_t}{dt}=C_nn_s(N_t-n_t)-{\beta}_nn_{s0}n_t                         
\end{equation} 

where $N_t$ is the concentration of the surface trap levels, $n_t$ is the concentration of captured 
electrons; $n_s$ is the concentration of the free charge carriers near the semiconductor 
surface and $C_n$ and $\beta_n$ are the probabilities for capturing and releasing of charge carriers. 
The concentrations $n_t$ and $n_s$ deviate from their equilibrium values $n_{t0}$ and $n_{s0}$ under the 
influence of ultrasonic waves:

\begin{equation}\label{8}
n_t=n_{t0}+{\Delta}n_t, \hspace{1cm} n_s=n_{s0}+{\Delta}n_s                         
\end{equation} 

The condition of thermodynamic balance in the absence of the acoustic wave is

\begin{equation}\label{9}
C_nn_{s0}(N_t-n_{t0})-\beta_nn_{s0}n_{t0}=0 
\end{equation} 
                                        
This equation provides the connection between the coefficients $C_n$ and $\beta_n$ . For a 
non-degenerate semiconductor we obtain:

\begin{equation}\label{10}
\beta_n=C_nexp((E_t-F_{s0})/kT)                         
\end{equation} 

where $F_{s0}$ is the Fermi level at the surface. By substituting equs. 8 and 10 into equ.7, we 
obtain:

\begin{equation}\label{11}
\frac{d{\Delta}n_t({\lambda})}{dt}=C_n{\lbrack}(N_t-n_{t0}({\lambda})){\Delta}n_s-
(n_{s0}({\lambda})+{\Delta}n_{s0}({\lambda})+N_ce^{-E_t/kT}){\Delta}n_t({\lambda}){\rbrack}                         
\end{equation} 

Note, that ${\Delta}n_t$ and ${\Delta}n_s$ are averaged over the acoustic wave period. 
Acoustoelectrical voltage component, which is connected to trapped charge, is 
proportional to the excess charge on said traps $(V_{ae}{\sim}{\Delta}n_t)$. Performing the 
measurement of the TAV signal and its shape under quasi-equilibrium 
conditions, it means:

\begin{equation}\label{12}
\frac{d{\Delta}n_t}{dt}=0                         
\end{equation} 

Then, from equ.11 we receive the following expression for the trap TAV 
component:

\begin{equation}\label{13}
V_{ae}\sim {\Delta}n_t({\lambda})=(N_t-n_{t0}({\lambda}))\frac{{\Delta}n_s({\lambda})}
{(n_{s0}({\lambda})+{\Delta}n_{s0}({\lambda})+N_ce^{-E_t/kT})}                         
\end{equation} 

The values $n_{t0}$, ${\Delta}n_s$ and ${\Delta}n_{s0}$ depend on sample illumination. However, if the 
photon energy $h\nu$ does not exceed the band gap energy $E_G$ and direct electronic excitation 
from valence band to conduction band does not occur, the changes of concentrations $n_{s0}$ 
and ${\Delta}n_{s0}$ are rather small. In this situation the TAV spectra is mainly determined by the 
first co-multiplier $(N_t-n_{t0}({\lambda}))$ in equ. 13. At  sample illumination with light of photon 
energies exceeding the value determined by equ.14:

\begin{equation}\label{14}
E_t=E_G-h\nu                         
\end{equation} 

the charge carriers transferring from the valence band to capture centers is 
resulting in a  growth of $n_{t0}$. Then the difference $(N_t-n_{t0}({\lambda}))$ in equ.13 decreases, and a 
minimum should be observed in the TAV spectrum. If several types of trap levels take 
part in TAV signal formation, then several corresponding minima should be observed in 
the optical TAV spectrum.

\begin{center}
\subsection{ANALYSIS OF THE RELAXATION TIME TEMPERATURE DEPENDENCY}
\end{center}

The value of an energy position of the deep trap levels $E_t$ can be obtained from 
temperature dependencies of the relaxation TAV time $\tau$ also. This dependency can be 
obtained from equ.3. The temperature dependency of $N_c$ and $V_T$ for nondegenerate 
semiconductor is well-known: $N_c\sim T^{3/2}$, $V_T\sim T^{1/2}$. For attracting Coulomb center at room 
temperature  $S_n\sim T^{-2}$ [11]. Therefore:

\begin{equation}\label{15}
N_c(T)V_T(T)S_n(T)\simeq const(T)                         
\end{equation} 

If this relation isn't absolute exact, the temperature dependence of expression (15) 
will much weaker than exponential dependence. Consequently, for deep levels $(E_t>>kT)$ 
and small temperature changes the temperature dependence of $(N_cV_TS_n)$ can be neglected 
in equ.(3). Having two times $\tau_1$ and $\tau_2$ for two different temperatures $T_1$ and $T_2$, we can to 
calculate $E_t$ by equ.(16):

\begin{equation}\label{16}
E_t=\frac{kT_2ln(\tau_1(T_1)/\tau_2(T_2))}{(T_2/T_1)-1}                         
\end{equation} 

One can enhance a precision of $E_t$ determination by using a plot of $ln(\tau)$ on $T^{-1}$, if the 
experiments are done in some temperature range:

\begin{equation}\label{17}
ln(\tau (T))=(E_t/kT)+const(T)                         
\end{equation} 

The curves of these dependencies are direct lines. The angles of declination of these 
plots are set by value $E_t$. Thus it is possible to define a value $E_t$ for each trap level.

\begin{center}
\section{SAMPLES AND EXPERIMENTAL TECHNIQUE}
\end{center}

We investigated three types of GaAs-samples. The first one labeled as GA-1 is the 
structures of epi-layer n-GaAs on n-GaAs substrate. They were 
fabricated by an industrial vapor phase epitaxy method in the system $Ga-AsCl_3-H_2$. The 
substrate, 0.35 mm thick, was doped by Te with $N_{Te}\approx (1\div2)\times10^{18} cm^{-3}$. The free carrier 
concentration of the epi-layers, doped by Te too, 6-9 microns thick for different samples, 
was $(0.6\div1.2)\times10^{15} cm^{-3}$. The second one labeled as GA-2 consists of a  $400 {\mu}m$ thick 
GaAs-substrate with a $1 {\mu}m$ thick epitaxial layer of n-GaAs on it. The third sample 
labeled as GA-3 consists of the GaAs substrate and a $8 {\mu}m$ thick epitaxial layer with 
electron concentration $n\sim 10^{14} cm^{-3}$. GA-2 and GA-3 samples were fabricated by 
MOCVD process. 

The samples GA-1 and GA-2 were placed by the epitaxial layer on the lithium 
niobate plate. On this plate  SAW of 6 MHz were excited. The TAV signal was picked 
up between a bottom ground electrode and a flat metal electrode attached on the upper 
surface of the GaAs sample as shown in Fig.2a. This arrangement is very convenient 
because no special sample preparation is required and sample replacement is easy to do. 
For example, it is not necessary to form electrical contacts to the samples. 
As far as GaAs is a piezoelectric material, it can simultaneously play the role of the 
piezoelectric waveguide as shown in Fig.2b (for GA-3). In this case SAW of 67 MHz are 
excited by means of an interdigital transducer deposited on the sample surface. The TAV 
signal was measured between the ground electrode and the aluminum film which has to 
be evaporated on the epitaxial layer. 

\begin{center}
\section{EXPERIMENTAL RESULTS AND DISCUSSION}
\end{center}

\begin{center}
\subsection{THE EXPERIMENTAL DEFINITION OF THE TRANSIENT TAV RELAXATION TIMES}
\end{center}

The task of the subsequent mathematical procedure consists in finding out of the 
values $V_i$ , $\tau_i$ and N from equ.6. To this aim it is necessary to determine the number N of 
effectively acting types of surface levels. Therefore we have to analyze the plots $ln(V_{ae})$ 
versus time. For  N=1 this dependence should be a straight line. Its slope gives the 
relaxation time $\tau$. In the case of two or more exponential components this plot has a more 
complex form. The number of rectilinear sites on these curve corresponds to the number 
of trap centers. One can define characteristic relaxation times from the declination angles 
of these directs. In our experiments the values $V_i$ and $\tau_i$ were estimated by the 
interpolation of the transient TAV signal (part "BC" in Fig.1) with the help of a special 
computer program which had found out the number of different type levels 
simultaneously. The data obtained from GA-2 show the occurrence of two exponential 
terms with decay times of  2 and 12 ms. In Fig. 3 the parts "ab" and "bc" of the curve 1 
correspond to two types of traps having the relaxation times 2 and 12 ms, respectively. 
The experiments with the GA-3 sample show two different terms in total TAV signal 
corresponding to the two  parts "ab" and "bc" of the curve 2 shown in Fig. 3. We can 
distinguish two types of trapping centers having relaxation times of 4.5 and 22 ms at 
room temperature. Four local centers having relaxation times of 2.2, 1, 22 and 1.7 ms 
were also found in the samples GA-1 at room temperature. 

Additional measurements were carried out for further identification these trap 
centers and for the definitions of their parameters. The relaxation time of a transient TAV 
was measured at various temperatures in samples GA-1  (see item 2.3). The samples 
GA-2 and GA-3 were used to measure spectral dependencies of a TAV signal (see item 
2.2).

\begin{center}
\subsection{TEMPERATURE DEPENDENCIES OF THE TRANSIENT TAV}
\end{center}

The descending parts of the TAV signal for GA-1 series at various temperatures are 
presented on Fig.4. We can see, that the TAV relaxation time decreases with temperature 
increase. Our measurements were carried out in a temperature range 294 to 330 K.

The plots of $\tau$ on $T^{-1}$ are shown on Fig.5. Practically they are direct lines, angles of 
declination of these lines give the trap level energy positions $E_t$. Knowing the 
characteristic relaxation times of the excess charge $\tau_i$ and the energy levels of the 
interface centers $E_{ti}$, we can calculate the effective capture cross section $S_{ni}$ for these levels 
by equ.(3). 

Thus, four deep levels were detected at the samples under study. They were 
designated as levels $E_{1,2,3,4}$. The characteristic parameters of said centers are presented in 
the Table. The characteristics coincidence of $E_4$ level and the literary data [12] allows to 
say that $E_4$ is electron center EL3. The configuration $As_iV_{Ga}$ is assigned with this level. 
Really, the excess of the interstitial arsenic $(As_i)$ and the gallium vacancies $(V_{Ga})$ should 
be observed in the GaAs, doped by tellurium. Certainly, the tellurium interstitial $(Te_i)$ can 
be significant at this case too. It is possible to conclude, that level $E_1$ is EL17 center [13-
15], and $E_3$ - EL5 center [14]. Level $E_2$ can be identified as electron center EL6 [16]. 
All these levels deals with the vacancies of gallium $(V_{Ga})$ and arsenic $(V_{As})$ [13-16]. 

\begin{center}
\subsection{EXPERIMENTAL SPECTRA OF THE TRANSIENT TAV}
\end{center}

Also, for determination of trapping level energy the optical spectra of transient 
TAV are investigated. A sample under investigation is illuminated with monochromatic 
light in the range  from 0.6 to 1.6 eV. Then TAV signal is separated for partial 
exponential components with different relaxation times $\tau_i$ with the help of a special 
computer program a general. Each such component is characterized by a certain value of 
partial amplitude $V_i$ which varies during illumination. Said amplitudes $V_i$ determines a 
contribution to a total TAV signal of i-type traps having relaxation time $\tau_i$. This 
contribution reaches a minimum at illumination of sample surface by monochromatic 
light with photon energy equal $E_G-E_{ti}$, in accordance with the formula (13). Thus a 
spectral position of a minimum in a spectrum of i-partial amplitude $V_i(h{\nu})$ undutiful 
determines an appropriate energetic level $E_{ti}$.

The spectra of partial amplitudes $V_i$ are shown on the Fig.6. The plots (1 and 2) are 
taken from GA-2 sample for relaxation times 12 and 2 ms, respectively. The minima on 
these curves correspond to two types of deep levels at $E_5^*$=0.48 eV and $E_4^*$=0.54 eV 
below a conduction zone.  The plots 3 and 4 are taken from GA-3 sample for relaxation 
times 22 ms and 4.5 ms, respectively. The minima on plots 3 and 4 correspond to two 
types of deep levels at $E_5$=0.48 eV and $E_1^*$=0.20 eV below a conduction zone.

Now, knowing the energy positions of trapping centers and the related relaxation 
times, we can calculate the effective cross sections $S_n$ for capture of charge carriers by 
these trap centers. Below, characteristic defect parameters in the epitaxial GaAs 
structures under study are presented in Table. The defect  identification was done by 
comparing the energetic position with literature data on deep levels in GaAs [12,13, 17-
20]. One can see, that deep trap levels of the types EL3 and EL17 are detected both  with 
the help of spectral researches and measurements of TAV relaxation times' temperature 
dependencies.

The difference in values of relaxation times for the levels $E_1$, $E_1^*$ and $E_5$ ,$E_5^*$ in 
different samples can be explained as follow. A relaxation time of captured charge is 
determined by two parameters - energy depth $E_t$ of trapping center and its capture cross-
section $S_n$. Said $E_t$ depends on atomic structure of local defect and so does not vary from 
a sample to sample. Opposite, the value $S_n$ depends both on a type of trapping level and 
on electrical properties of semiconductor itself, including a configuration of electrical 
potential near defect and its shielding by a free charge. Therefore the characteristic 
relaxation times for the same trapping levels, but in various on manufacturing samples, 
can differ from each other.

\begin{center}
\section{CONCLUSIONS}
\end{center}

The acoustoelectric method can be advantageously applied for studying interfaces 
in modern epilayer semiconductor structures. The experimental data followed by 
appropriate mathematical processing allow to estimate characteristic relaxation time, 
energy position in the forbidden zone and electron capture cross section for each of these 
centers. By this approach we, for the first time, measured the relaxation times of trap 
centers.

Investigating the interface between epitaxial GaAs-layers grown on substrate 
by MOCVD process and chlorine vapor method we found five different types of trapping 
defects. They were identified as EL3, EL4, EL5, EL6 and EL17 centers. Their 
characteristics at room temperature are summarized in the Table. 

The different GaAs centers are known to be observed in the samples, fabricated by 
different technique [13]. The coincidence of the level's parameters permit to conclude 
that levels EL5 and EL17 can be present in the epitaxial structures, fabricated by chlorine 
vapor method as well as by MOCVD technique.

The method described can also be applied to the investigation of other 
semiconductor structures, including those based on Si and $A_2B_6$ compounds.

Summarizing the obtained results we may conclude that the transient TAV 
technique is a reasonably simple and effective method for characterizing surface and 
interface trap centers.

\begin{center}
\section*{ACKNOWLEDGMENTS}
\end{center}

We kindly acknowledge the Bundesministerium f\"ur Bildung, Wissenschaft, 
Forschung und Technologie, Deutschland, and the ISSEP Program of International Soros 
Foundation for partial financial supporting of the project.

\newpage
\begin{center}
\section*{REFERENCES}
\end{center}

[1]  V.A. Jurjev, V.P. Kalinushkin and O.V. Astafjev. Phys. and techn. of semicon., {\bf29}, 
455 (1995).

[2]  A.S. Popov and A.Y. Bahnev. Phys. Stat. Sol. (a), {\bf122}, 569 (1990).

[3]  Yu.D. Tkachev, V.S. Lysenko and V.I. Turchanikov. Phys. Stat. Sol. (a), {\bf140}, 163 
(1993).

[4]  A. Abbate, ë. H n, I. Ostrovskii and P. Das. Solid St. Electron., {\bf36}, 697 (1993).

[5]  I.V. Ostrovskii, A. Abbate, K.J. Han and P. Das. IEEE transition on ultrasonic, 
ferroelectric, and frequency control Conf. Proc. {\bf42}, 876 (1995).

[6]  I.V. Ostrovskii and S.V. Saiko. Phys. and techn. of semicon., {\bf28}, 796 (1994).

[7]  M. Tabib-Azar, Nam-Chun Park and P. Das. Solid St. Electron., {\bf30}, 705 (1987).

[8]  V.A. Vyun, V.V. Pnev and I.B. Yakovkin.. Surface Waves in Solids and Layered 
Structures, Novosibirsk Conf. Proc.  {\bf2}, 354 (1986).

[9]  Ju.V. Guljaev, A.M. Kmita, A.V. Medved, V.P. Plesskii, N.N. Shybanova and V.N. 
Fedorez. Sov. phys. stat. sol., {\bf17}, 3505 (1975).

[10]  A. Ricksand and O. Engstrom. J. Appl. Phys., {\bf70}, 6915 (1991).

[11]  V.N. Abakumov, I.N. Yassievitch, JETP, {\bf71}, 657 (1976).

[12]  M.M. Sobolev et al. Semiconductors, {\bf30}, 1108 (1996).

[13]  G.M. Martin, A. Mitoneau, A. Mircea. Electron. Lett., {\bf13}, 172 (1977).

[14]  V.A. Samojlov, N.Ja. Jakhusheva, V.Ja. Princ. Semiconductors, {\bf28}, 1617 (1994).

[15]  R.A. Morrou. J. Appl. Phys., {\bf69}, 3396 (1991).

[16]  Zh.-O. Fang, Schlesinger. J. Appl. Phys., {\bf61}, 5047 (1987).
 
[17]  E.A. Bobrova, G.N. Calkin, V.L. Oplesnin, M.G. Tigishvili. Surface. Physics, 
Chemistry and Mechanics., {\bf3}, 130 (1991). 

[18]  A.P. Kasatkin, V.A. Perevoshchikov, V.D. Skupov, L.A. Suslov. Surface. Physics, 
Chemistry and Mechanics, {\bf6}, 79 (1993). 

[19]  B.I. Siosoev, N.K. Bezredin, G.I. Kotov. Sov. Phys. stat. sol., {\bf29}, 24 (1995).

\newpage
\begin{center}
\section*{Table Captions}
\end{center}

Table. Characteristics of interface trapping centers in epi-GaAs at room temperature. 

\begin{center}

\begin{tabular}{|c|c|c|c|c|c|}
\hline
 Samples & Level & Type & $E_t$, eV & $\tau$, ms & $S_n$, $cm^2$ \\
\hline
GA-1&$E_1$&EL17&$0.23\pm0.02$&$2.2\pm0.2$&$0.7\times10^{-18}$\\
\hline
GA-3&$E_1^\ast$&EL17&$0.20\pm0.01$&$4.5\pm0.4$&$1.5\times10^{-19}$\\
\hline
GA-1&$E_2$&EL6&$0.29\pm0.02$&$1.0\pm0.1$&$1.6\times10^{-17}$\\
\hline
GA-1&$E_3$&EL5&$0.42\pm0.02$&$22\pm2$&$1.4\times10^{-16}$\\
\hline
GA-1&$E_4$&EL3&$0.56\pm0.02$&$1.7\pm0.2$&$4\times10^{-13}$\\
\hline
GA-2&$E_4^\ast$&EL3&$0.54\pm0.01$&$2\pm0.03$&$2.5\times10^{-13}$\\
\hline
GA-3&$E_5$&EL4&$0.48\pm0.01$&$22\pm1$&$2\times10^{-15}$\\
\hline
GA-2&$E_5^\ast$&EL4&$0.48\pm0.01$&$12\pm1$&$4\times10^{-15}$\\
\hline
\end{tabular}
\end{center}

\newpage
\begin{center}
\section*{Figure Captions}
\end{center}

Fig.1.  Time dependency of input rf-voltage V exciting SAW and resulting TAV signal.
 
Fig.2.  Two versions of sample arrangements performing transient TAV measurements:
a) separate medium configuration, b) integrated configuration

Fig.3.  Logarithmic plot of transient TAV signal versus time. 1 is plot for GA-2 and 2 is 
plot for GA-3. Parts "ab",  and "bc" correspond to different  types of trapping 
centers.

Fig.4.  Shape of the descending parts of the TAV signal at various temperatures for GA-
1: 296 K (1), 305 K (2) and 318 K (3). Points - experimental data; curves - 
theoretical calculation by equ.(6)

Fig.5.  Dependency of $\tau$ on $T^{-1}$ (GA-1). At room temperature the relaxation time $\tau$ is 
equal to 2.2 ms for the plot 1, 1.0 ms plot 2, 22 ms plot 3 and 1.7 ms plot 4. The 
slops of plots 1, 2, 3 and 4 give four levels $E_1$, $E_2$, $E_3$, $E_4$, respectively.

Fig.6.  Spectra of the TAV exponential component. 1- $\tau$ = 12 ms (GA-1); 2- $\tau$ = 2 ms 
(GA-2); 3- $\tau$ = 22 ms (GA-3); 4- $\tau$ = 4.5 ms (GA-2). The four minima labeled as 
$E_1^*$, $E_4^*$, $E_5$, $E_5^*$ correspond to the three trapping centers detected in the samples 
under study. The band gap energy is marked by $E_G$.

\end{document}